\documentclass{PoS}

\title{Quasi-PDFs from Twisted mass fermions at the physical point}

\ShortTitle{Quasi-PDFs from Twisted mass fermions at the physical point}

\author{Constantia Alexandrou$^{ab}$,
  Krzysztof Cichy$^{c}$,
        Martha Constantinou$^{d}$,  
	Kyriakos Hadjiyiannakou$^{b}$, 
	Karl Jansen$^{e}$,
	 \speaker{Aurora Scapellato}$\;^{a,f}$,
	Fernanda Steffens$^{g}$
\\
\\
        \llap{$^a$}Department of Physics, University of Cyprus, P.O. Box 20537, 1678 Nicosia, Cyprus\\
	\llap{$^b$}Computation-based Science and Technology
	Research Center, Cyprus Institute, 20 Kavafi Str.,
	Nicosia 2121, Cyprus\\
	\llap{$^c$}Faculty of Physics, Adam Mickiewicz
	University, Umultowska 85, 61-614 Pozna\'{n}, Poland\\
	\llap{$^d$}Temple University, 1925 N. 12th Street,
	Philadelphia, PA 19122, USA\\
	\llap{$^e$}John von Neumann Institute for Computing (NIC), DESY, Platanenallee 6, D-15738 Zeuthen, Germany\\
	\llap{$^f$}University of Wuppertal, Gau\ss str. 20, 42119 Wuppertal, Germany\\
	\llap{$^g$}Institut f\"{u}r Strahlen- und Kernphysik, Rheinische Friedrich-Wilhelms-Universit\"{a}t Bonn, Nussallee 14-16, 53115 Bonn\\
        E-mail: \email{scapellato.aurora@ucy.ac.cy}}

\abstract{We present results for the flavor non-singlet $u-d$ parton distribution functions within the nucleon  using the quasi-PDF approach. The lattice calculation is performed by employing the twisted mass formulation and two gauge ensembles, having $N_f=2$ and $N_f=2+1+1$ dynamical fermions with masses tuned to their physical value. For the $N_f=2$ physical point ensemble, the  unpolarized, helicity and transversity distributions are computed for three values of the nucleon momentum, namely $[6,8,10]\pi/L$ corresponding to  [0.83,1.11,1.38] GeV. Upon renormalization, we find that, as the nucleon momentum increases, the lattice results  approach the phenomenological distributions resulting from analyses of deep inelastic scattering data, opening a promising path for a direct evaluation of parton distributions from the QCD Lagrangian.
For the  $N_f=2+1+1$ physical point ensemble, we present preliminary results for the unpolarized distribution extracted from a nucleon boosted by $8\pi/L$ or 0.97 GeV.}

\FullConference{The 36th Annual International Symposium on Lattice Field Theory - LATTICE2018\\
		22-28 July, 2018\\
		Michigan State University, East Lansing, Michigan, USA.}

\let\OLDthebibliography\thebibliography
\renewcommand\thebibliography[1]{
  \OLDthebibliography{#1}
  \setlength{\itemsep}{-2pt}
}

\begin{document}

\section{Introduction}
\noindent Hadrons are complex systems whose internal structure is determined by the strong interactions among quarks and gluons, known as \textit{partons}. An open question in particle physics is how the decomposition of the hadron momentum among its constituents emerges from Quantum Chromodynamics (QCD). In this framework, one of the most essential quantities are the parton distribution functions (PDFs), which describe the probability density that a parton carries a collinear momentum fraction $x$ of the total momentum of the hadron. PDFs are intrinsically non-perturbative objects and require light-cone dynamics, which however cannot be accomodated on a Euclidean lattice, where a finite lattice spacing is employed. For this reason, global QCD analyses of deep inelastic scattering data, aided by phenomenological models, have been the only possible source of information up to date~\cite{Lin:2017snn}.
In this work, we report our effort in extracting  the $x$-dependence of the isovector ($u-d$) distribution functions for unpolarized and polarized nucleons by using the quasi-PDF approach~\cite{Ji:2014gla}. Other proposals to access PDFs from first principles have also been recently suggested, like pseudo-PDFs~\cite{Radyushkin:2017cyf} implemented within lattice QCD in Ref.~\cite{Orginos:2017kos} and the construction of good Lattice Cross Sections (LCSs)~\cite{Ma:2017pxb}. Here we show our results for unpolarized, helicity and transversity PDFs extracted from $N_f=2$ physical point simulations, also presented in~\cite{Alexandrou:2018pbm,Alexandrou:2018eet}, and report preliminary results for unpolarized PDFs obtained analyzing a $N_f=2+1+1$ ensemble, where all quark masses are tuned to their physical value. The simulation parameters of the ensembles are listed in Table~\ref{Table:parameters_ensembles}.
{\small{
\begin{table}[h]
\centering
\renewcommand{\arraystretch}{1.1}
\renewcommand{\tabcolsep}{5pt}
\begin{tabular}{c c c c c c c c c}
 \hline\hline
$N_f$ & $T/a, L/a$ & $\beta$ & $c_{sw}$ & $\kappa$ & $a\mu$ & $aM_{\pi}$ & $a$[fm] & $M_{\pi}L$ \\
 \hline
2 & 96, 48 & 2.10 & 1.57751 & 0.137290 & 0.0009 & 0.06208(2) & 0.0938(2) & 2.98\\ 
 \hline
2+1+1 & 128, 64 & 1.778 & 1.69 & 0.1394265 & 0.00072 & 0.05658(6) & 0.08029(41) & 3.62\\ 
 \hline
\end{tabular}
\caption{Simulation parameters of the $N_f=2$ ensemble~\cite{Abdel-Rehim:2015pwa} and $N_f=2+1+1$ ensemble~\cite{Alexandrou:2018egz} employed in this work.}
\label{Table:parameters_ensembles}
\end{table}
}}
\vspace*{-0.5cm}
\section{Results from $N_f=2$ physical point simulations}

\begin{figure}[h!]
\centering
\includegraphics[width=.47\textwidth]{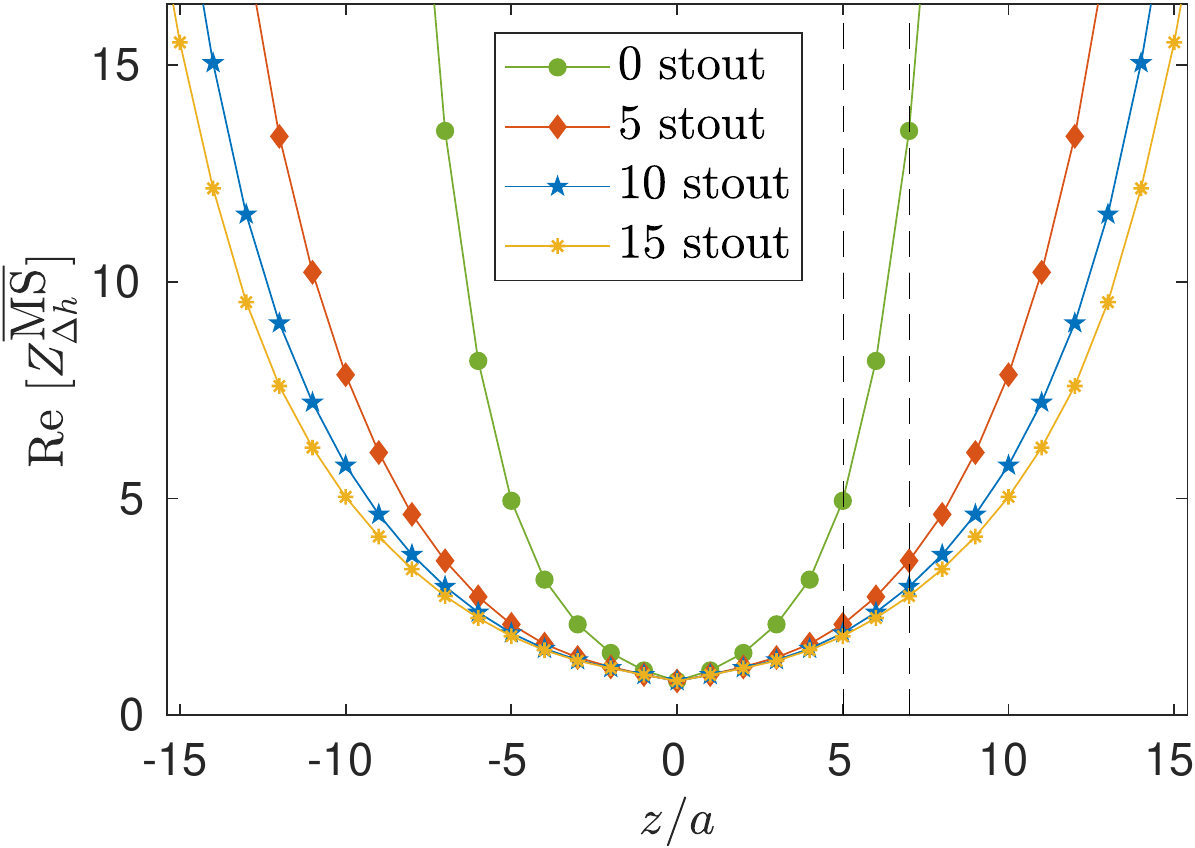}
\hspace*{0.6cm}
\includegraphics[width=.477\textwidth]{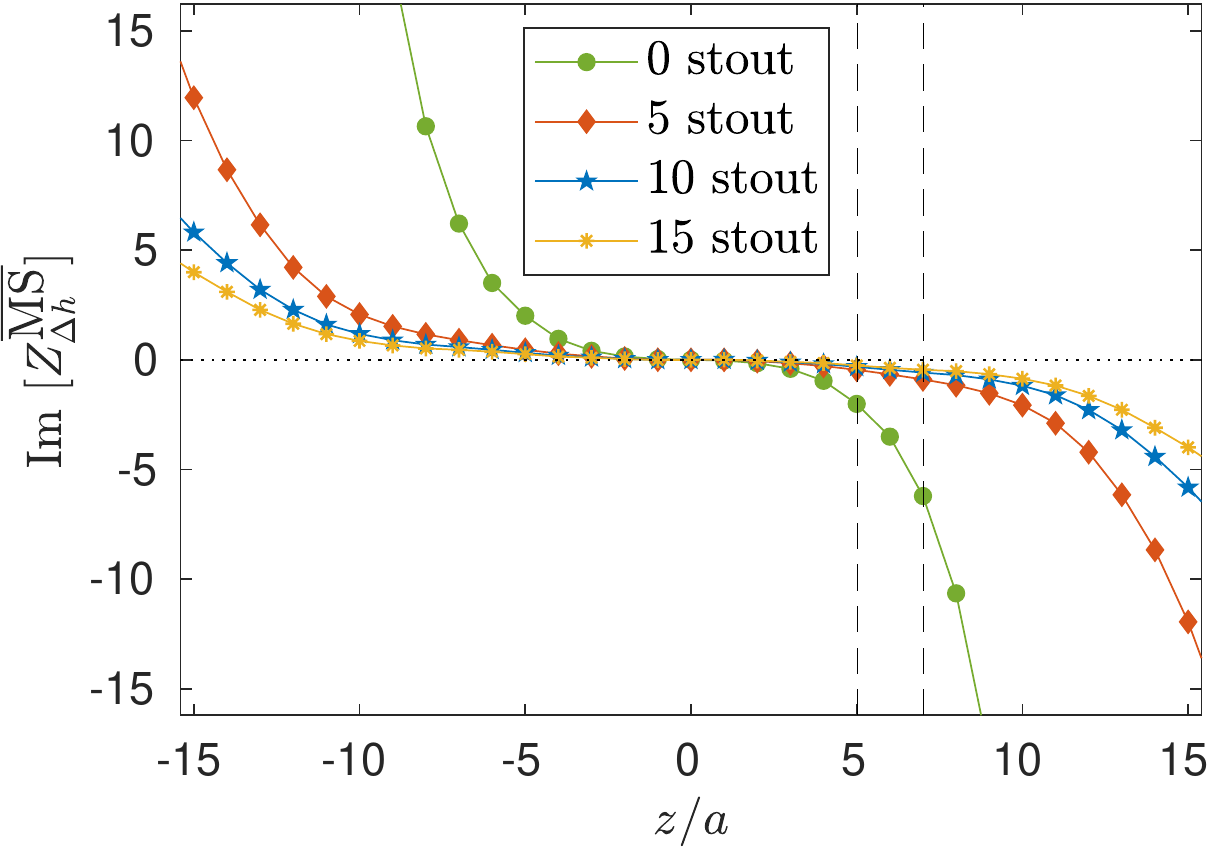}
\caption{Real (left) and imaginary (right) part of the Z-factors in ${\overline{\mbox{MS}}}$ at 2 GeV for the helicity operator, applying $N_{st}=0, 5, 10, 15$ and 30  levels of stout smearing to the Wilson line.}
\label{fig:Z_48c}
\end{figure}

\begin{figure}[h!]
\centering
\includegraphics[width=.47\textwidth]{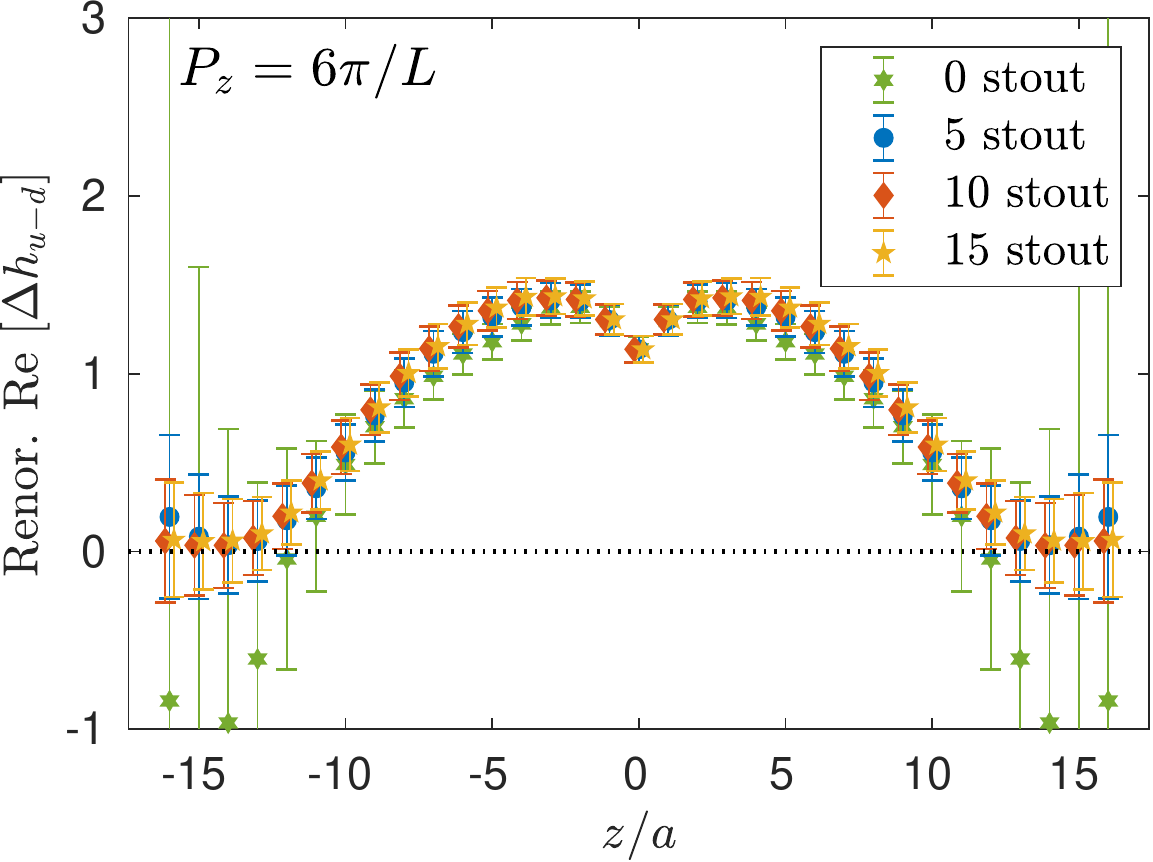}
\hspace*{0.6cm}
\includegraphics[width=.47\textwidth]{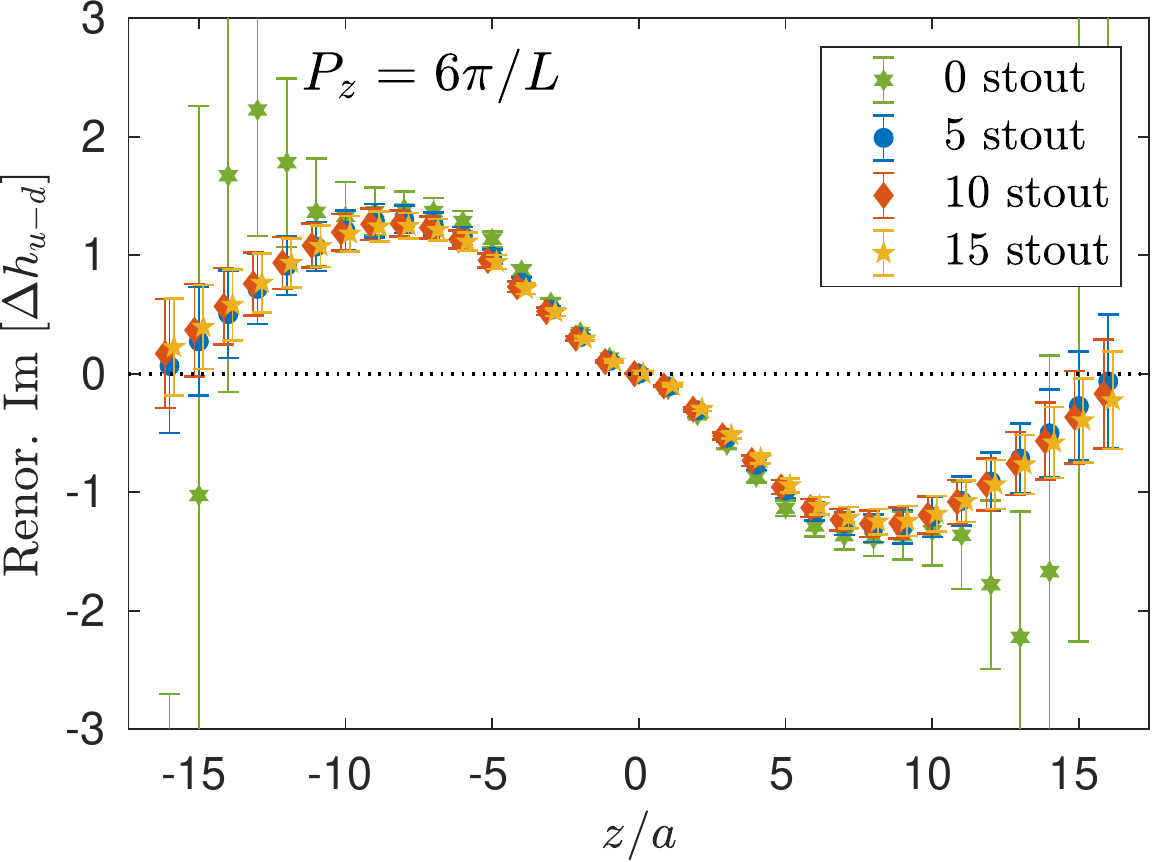}
\caption{Real (left) and imaginary (right) part of the renormalized matrix elements for the unpolarized PDF, extracted from a nucleon boosted by $6\pi/L$ and source-sink separation set to 1.12 fm.}
\label{fig:ME_renorm_48c}
\end{figure}

\noindent For the $N_f=2$ ensemble we compute the matrix elements relevant for quasi-PDFs at three nucleon momenta, $6\pi/L$, $8\pi/L$ and $10\pi/L$ or 0.83, 1.11 and 1.38 GeV, using sequential inversions through the sink and the momentum smearing technique~\cite{Bali:2016lva}. More details about the setup and statistics can be found in Refs.~\cite{Alexandrou:2018pbm,Alexandrou:2018eet} and~\cite{Krzysztof_proc:18}, where the effects of excited states on the bare matrix elements are also discussed. The results presented in the following are obtained from our largest source-sink time separation, namely $T_{sink}=12a\simeq 1.12$~fm, where ground state convergence is observed  within our statistical uncertainties, of around $10\%$. 
Having computed the lattice  matrix elements  (see~\cite{Alexandrou:2018pbm,Alexandrou:2018eet, Krzysztof_proc:18}), we renormalize them by adopting a RI$'$-type scheme.
The renormalization functions (Z-factors) are extracted by imposing the renormalization conditions described in Ref.~\cite{Alexandrou:2017huk}. We study the RI$'$-scale dependence on the vertex functions exploring 17 choices for the RI$'$ scales, having the form $a\bar{p}=\frac{2\pi}{48}\left(\frac{n_t}{2}+\frac{1}{4},n,n,n\right)$ with $\hat{P}=\frac{\sum_i \bar{p}_i^4}{\left(\sum_i \bar{p}_i^2\right)^2}\sim 0.25$ in order to
suppress discretization effects~\cite{Alexandrou:2015sea}. We take as final choices the ones for which $(a\bar{p})\in [1.0-2.6]$ to avoid contamination from non-perturbative effects. The
Z-factors are then converted from RI$'$ to the
$\overline{\mbox{MS}}$ scheme and evolved to $\bar{\mu}=2$ GeV using the 1-loop formulae of Ref.~\cite{Constantinou:2017sej}. To remove the residual dependence on $(a\bar{p})^2$, we perform a linear fit in $(a\bar{p})^2$ for each length $z$ of the Wilson line, of the kind $Z^{\overline{\mbox{\small{MS}}}}
=Z_0^{\overline{\mbox{\small{MS}}}}+(a\bar{p})^2 Z_1^{\overline{\mbox{\small{MS}}}}$, and we take  $Z_0^{\overline{\mbox{\small{MS}}}}$ 
as final Z-factor. In this work, we also study how the vertex functions depend on the smearing applied to the gauge field, as the smearing mitigates the power  divergences emerging for non-local operators.
In our lattice computation, we apply 3-dimensional stout smearing
only to the Wilson line connecting the fermion fields and the result for the Z-factors is shown in Fig.\ref{fig:Z_48c} for the helicity operator.
As can be seen, smearing clearly reduces the value of Z-factors for $z/a>3$ and the change from $N_{st}=0$ to $N_{st}=5$ is  more significant than from 5 to 10. In order to renormalize the matrix elements properly, the bare matrix elements have to be produced by applying the same number of iterations of stout smearing. The smearing dependence is considerable at the level of the bare matrix elements~\cite{Alexandrou:2017dzj}, but the renormalized matrix elements agree as expected and demonstrated in Fig.\ref{fig:ME_renorm_48c}.

\begin{figure}[h!]
\centering
\includegraphics[width=.47\textwidth]{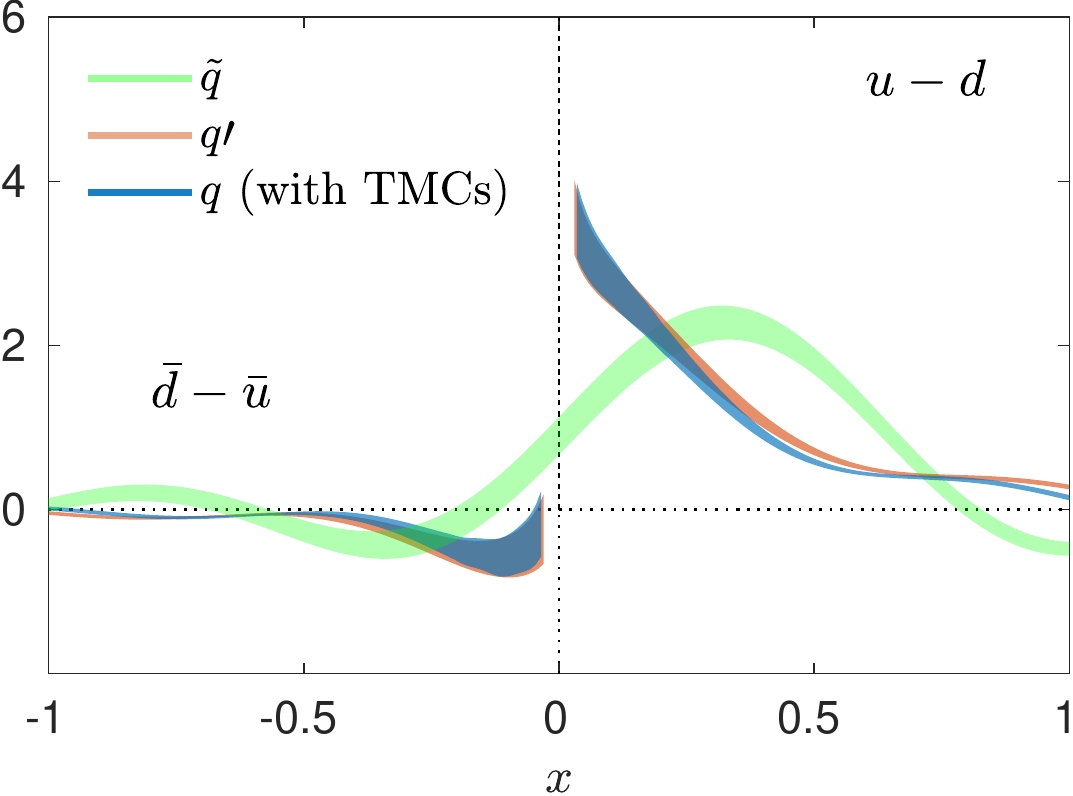}
\hspace*{0.65cm}
\includegraphics[width=.47\textwidth]{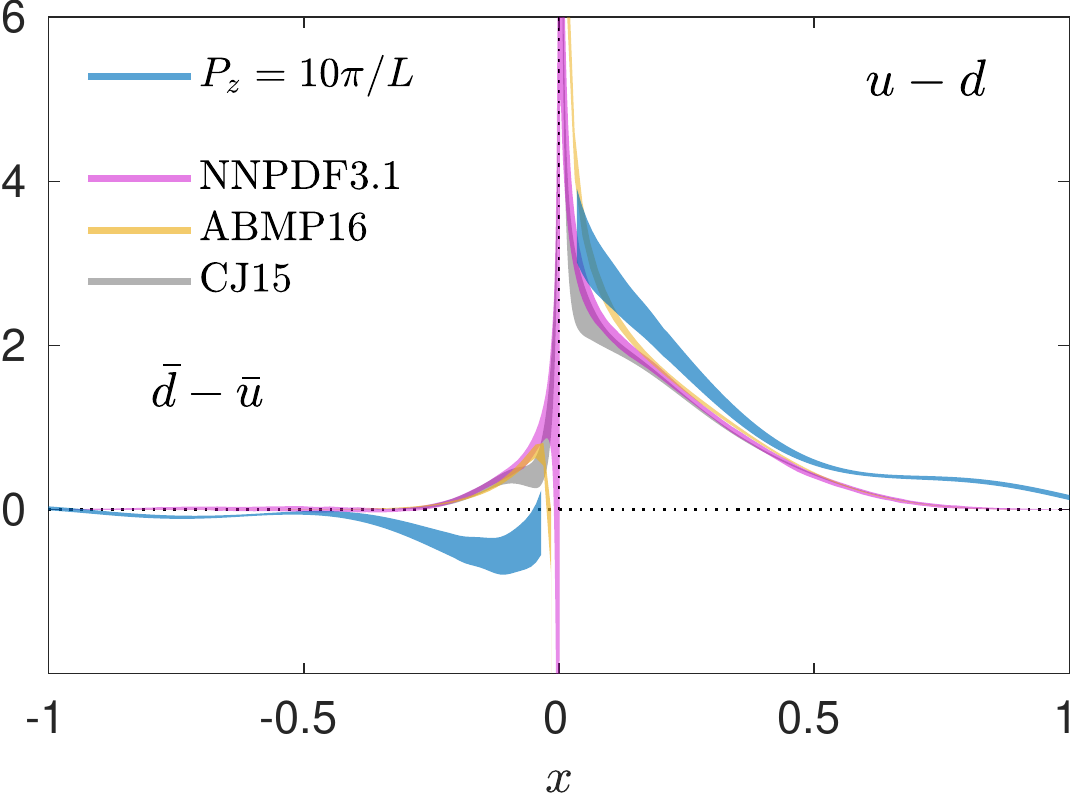}
\caption{Unpolarized distribution at $10\pi/L$. Left: Quasi-PDF (green), matched distribution (red), final PDF, after TMCs are applied (blue). Right: comparison with NNPDF3.1~\cite{Ball:2017nwa}, ABMP16~\cite{Alekhin:2017kpj}, CJ15~\cite{Accardi:2016qay}.}
\label{fig:matched_48c}
\end{figure}

\noindent In the quasi-PDF approach, the first step towards the reconstruction of the physical PDFs is the Fourier transform of the renormalized matrix elements. The renormalized Fourier transform, known as \textit{quasi-PDF}, is shown in Fig.\ref{fig:matched_48c} (left plot) by the green curve, for nucleon momentum $10\pi/L$. The quasi-PDF does not represent yet a physical result, until  a matching procedure and target mass corrections (TMCs) are applied to account for a finite value of the momentum employed on the lattice. In this work, we use the perturbative matching formulae of Refs.~\cite{Alexandrou:2018pbm, Alexandrou:2018eet} that we developed for the unpolarized, helicity and transversity PDFs, and for the TMCs we use Ref.~\cite{Chen:2016utp}. The matching procedure drastically alters the slope of the quasi-PDF in the whole x-range, as can be seen from the red band in Fig.\ref{fig:matched_48c} (left plot). It shifts and enhances the peak towards the small x-region, reproducing the antisymmetry between quark-antiquark distributions expected from global QCD analyses. TMCs are found to have a small effect, mostly evident in the positive $x$-region.
\vspace*{-0.3cm}
\begin{figure}[h!]
\centering
\includegraphics[width=.47\textwidth]{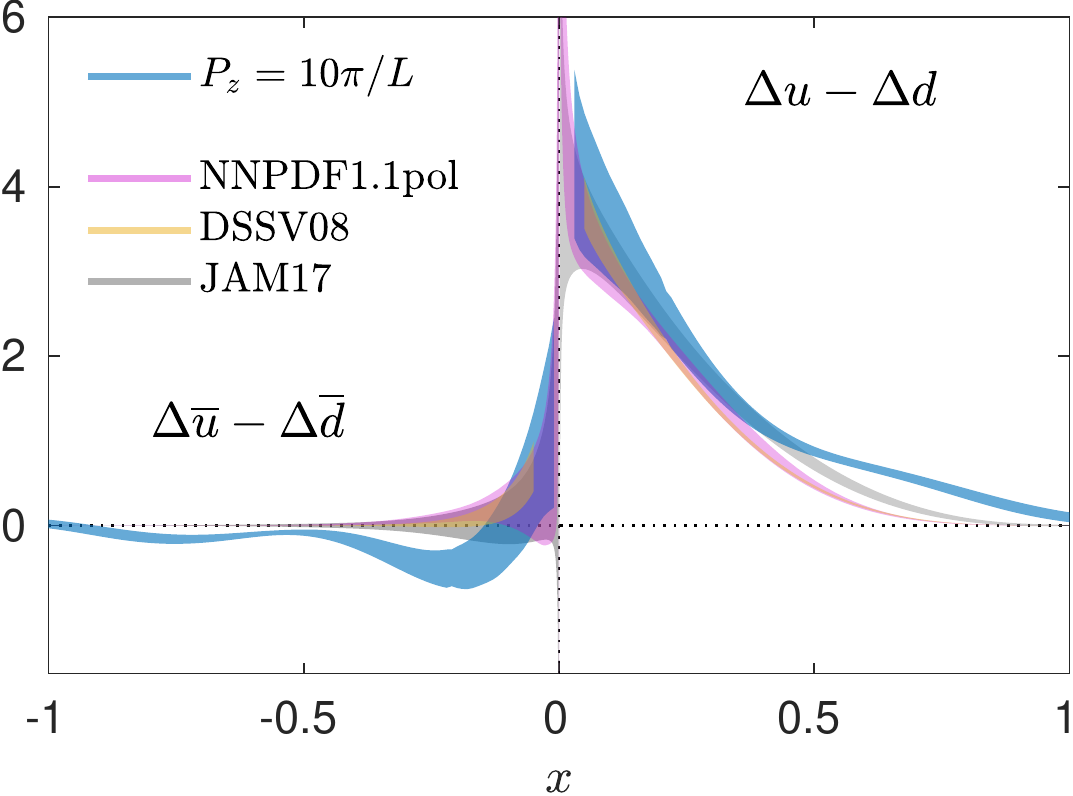}
\hspace*{0.66cm}
\includegraphics[width=.47\textwidth]{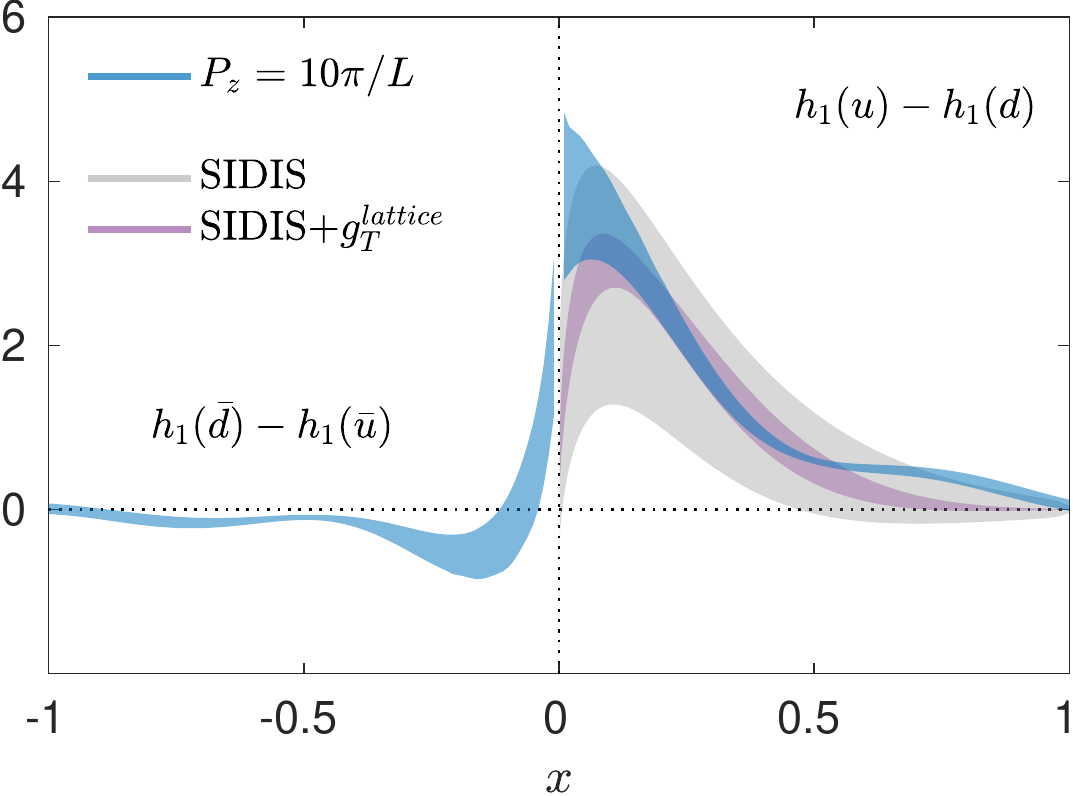}
\caption{Left: Helicity PDF from this work (blue curve) compared with NNPDF1.1pol~\cite{Nocera:2014gqa}, DSSV08~\cite{deFlorian:2009vb} and JAM17~\cite{Ethier:2017zbq} data. Right: Comparison of transversity PDF at boost $10\pi/L$ (blue curve), SIDIS data (grey band)~\cite{Lin:2017stx} and SIDIS data constrained with $g_T$ extracted from lattice QCD~\cite{Lin:2017stx} (violet band).}
\label{fig:polarized_48c}
\end{figure}

\noindent After applying the matching and TMCs to the quasi-PDFs computed at different momenta, one is ready to check how the lattice PDFs depend on the value of the boost and how they compare with phenomenology. As concerns the momentum dependence we refer to Refs.~\cite{Alexandrou:2018pbm,Alexandrou:2018eet}, where we find that, as the momentum increases, lattice PDFs move towards the phenomenological PDFs, a feature that enhances our confidence in the quasi-PDF approach. Here we show our results for the largest boost employed in this work. For the unpolarized PDF (right plot in Fig.\ref{fig:matched_48c}) we observe a similar slope as compared with global QCD analyses, while for the helicity we obtain agreement with phenomenology for $0<x<0.4$ and for small negative $x$, as can be seen in the left plot of Fig.\ref{fig:polarized_48c}. 
Thus, a fully reliable reconstruction of PDFs from the lattice can become possible for even higher momenta, maintaining control over a number of systematic effects~\cite{Krzysztof_proc:18}, such as contamination by excited
states and cut-off effects.
For the transversity PDF shown in Fig.\ref{fig:polarized_48c}, we find that the statistical errors
of the lattice PDFs are strikingly smaller than the phenomenological fits when use only SIDIS data and this also holds for the SIDIS data constrained with the tensor charge $g_T$ computed from lattice. 

\vspace*{-0.1cm}
\section{Quasi-PDFs from $N_f=2+1+1$ physical point simulations}
\begin{figure}[h!]
\centering
\includegraphics[width=.47\textwidth]{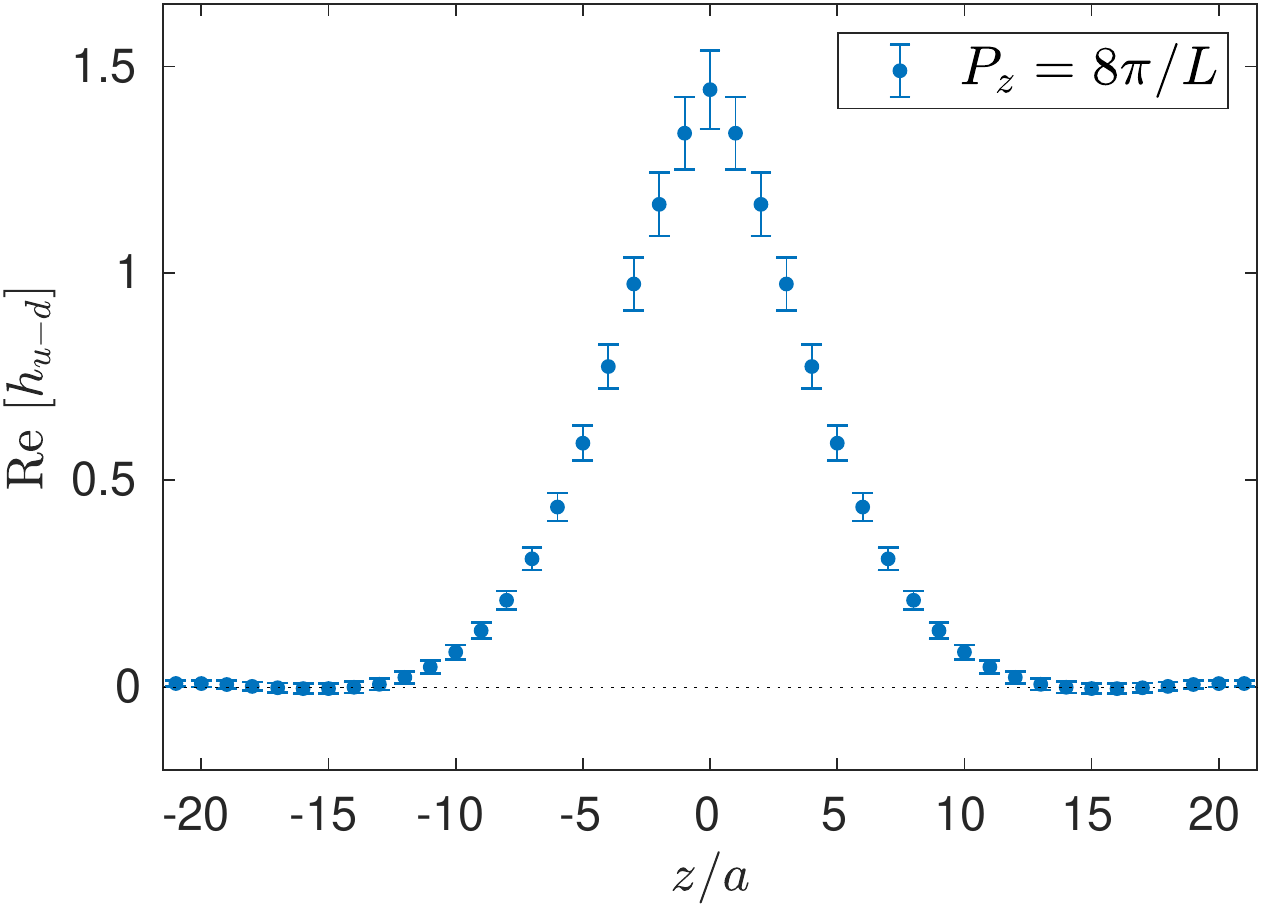}
\hspace*{0.5cm}
\includegraphics[width=.474\textwidth]{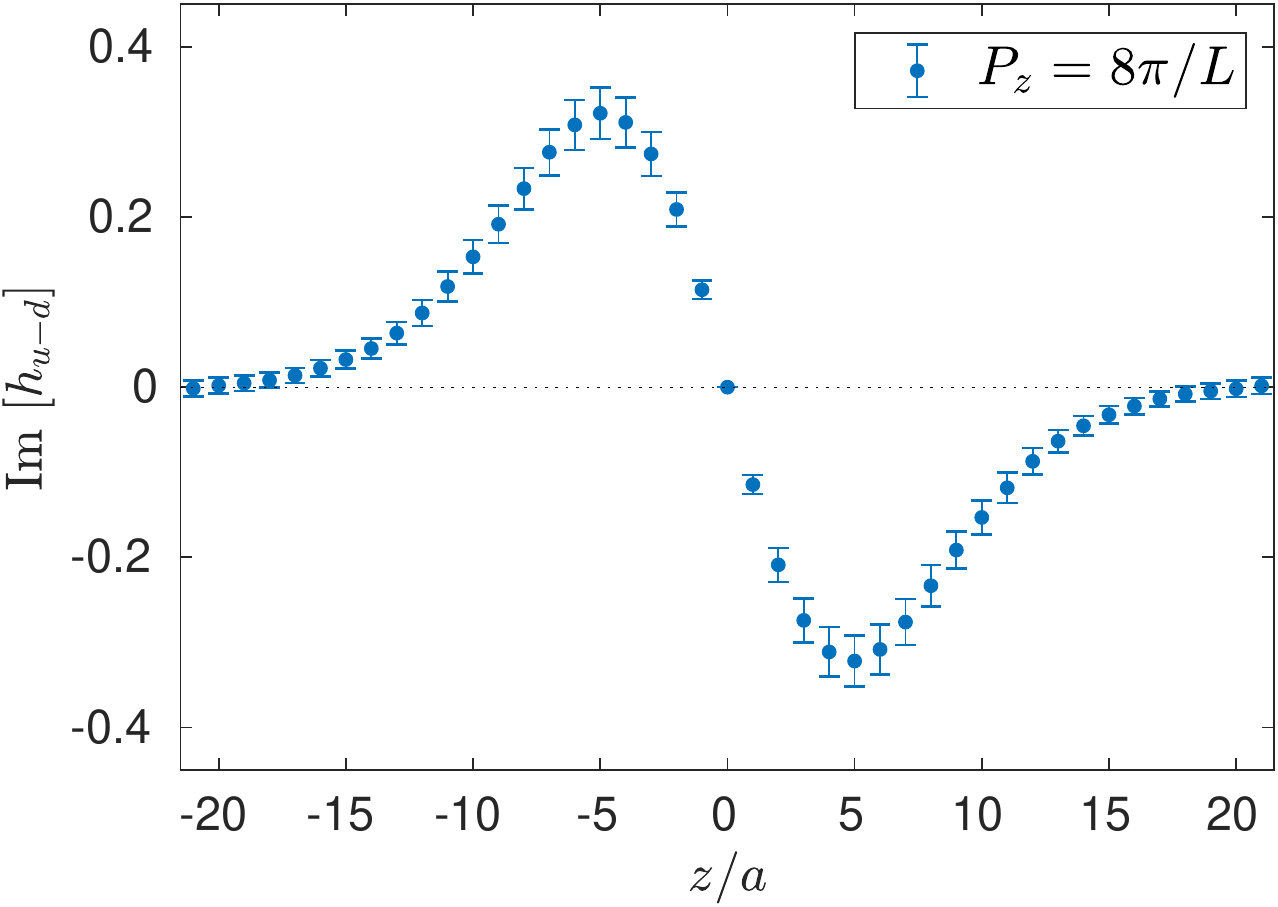}
\caption{Real (left) and imaginary (right) part of the unrenormalized matrix elements extracted using the $\gamma_0$ for the unpolarized PDF, for nucleon boosted by $8\pi/L$ and source-sink separation set to 1.12 fm.}
\label{fig:ME_64c}
\end{figure}

\begin{figure}
  \begin{minipage}{0.42\linewidth}
   \center{\includegraphics[width=1.04\linewidth]{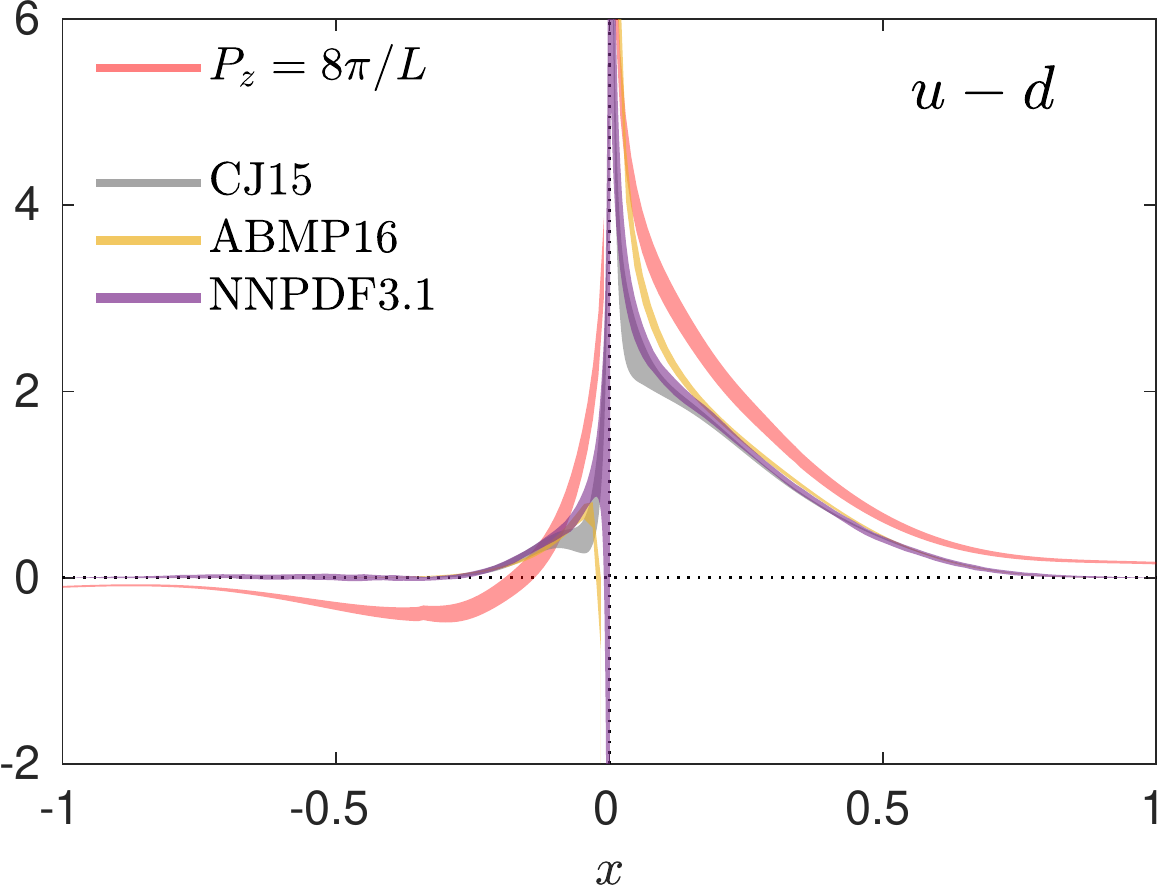}}
  \end{minipage}
  \hspace{1.5cm}
 \begin{minipage}{0.44\linewidth}
  \caption{Unpolarized PDF extracted from $N_f=2+1+1$ lattice QCD simulations at boost $8\pi/L$ (red band). The lattice PDF is compared to three phenomenological data from CJ15~\cite{Accardi:2016qay}, ABMP16~\cite{Alekhin:2017kpj}, NNPDF3.1~\cite{Ball:2017nwa} (grey, yellow, violet curves).}
  \label{fig:unpolarized_64c}
  \end{minipage} 
\end{figure}

\vspace*{-0.1cm}
\noindent As a next step towards a better control of systematic effects, we compute PDFs on a physical point ensemble simulated with mass-degenerate \textit{up} and \textit{down} quarks and a \textit{strange} and \textit{charm} quark in the sea, since eventually our aim is to take the continuum limit of hadron observables using as complete description of QCD as possible. 
We employ gauge configurations recently generated by the ETM collaboration~\cite{Alexandrou:2018egz} with parameters listed in Table~\ref{Table:parameters_ensembles}.
\noindent This ensemble, as compared to the $N_f=2$ previously analyzed by us, has a smaller lattice spacing, bigger volume and a larger value for $M_\pi L$. Thus, we expect discretization and finite volume effects to be less significant for this setup. 
The lattice computation of PDFs proceeds along the lines of the one performed on the $N_f=2$ ensemble and uses the same optimized setup of Ref.~\cite{Alexandrou:2018egz}, including momentum smearing.  
In Fig.\ref{fig:ME_64c} we show the bare matrix elements for the unpolarized PDF obtained from the analysis of $50$ configurations with total statistics of 4500 measurements, at $T_{sink}=14a\simeq 1.12$~fm and 
nucleon boost $P_z=8\pi/L\simeq 0.97$~GeV. 
For the renormalization we use, at this preliminary stage, one RI$'$ scale which is set to $a\bar{p}=\frac{2\pi}{64}\left(\frac{13}{2}+\frac{1}{4},8,8,8\right)$, with $(a\bar{p})^2 >1$ and $\hat{P}\sim 0.25$.
A detailed study on the RI$'$  scale dependence of the renormalization factors will be carried out in the near future. Having then converted the Z-factors to $\overline{\mbox{MS}}$ at 2 GeV and applied the matching kernel of Ref.~\cite{Alexandrou:2018pbm} and the TMCs of Ref.~\cite{Chen:2016utp}, we obtain the result depicted  in Fig.\ref{fig:unpolarized_64c}. A nice feature is that even at the value of the nucleon boost employed here, the lattice results exhibit a similar slope as PDFs from global analyses for $x>0$ and approaches zero at $x=-1$. The disagreement with phenomenological data is however expected for a number of reasons, most notably a small value of the nucleon boost, for which higher twist effects may play an important role in the matching of the quasi  PDFs. The calculation at higher boosts is one of our essential directions in pursuing the extraction of quark distributions from this $N_f=2+1+1$ physical point ensemble.
\vspace*{-0.1cm}
\section{Conclusions and outlook}
\vspace*{-0.1cm}
\noindent In this work we report our results on the extraction of PDFs from lattice QCD simulations performed with $N_f=2$ and $N_f=2+1+1$ ensembles at the physical pion mass. Our results have several nice features, the most important of which is the tendency of the lattice PDFs to approach the phenomenological ones as the nucleon boost increases. This is  motivating further studies that may improve the lattice estimates on several points, such as the subtraction of discretization effects in the renormalization functions and the calculation of the conversion factor RI$'$ to $\overline{\mbox{MS}}$ up to 2-loops. The computation at higher momentum is currently ongoing. The investigation of finite volume effects and cut-off effects at the physical value of quark masses is also planned within our  long-term program. Availability of large computational resources to study these systematic effects would be needed for a reliable check.
\vspace*{-0.3cm}
\section{Acknowledgment}
\vspace*{-0.1cm}
\noindent This  work  has  received  funding  from
the  European  Union's  Horizon  2020  research  and  innovation programme under the Marie Sk$\l$odowska-Curie
grant agreement No 642069 (HPC-LEAP). K.C. is supported  by  the  National  Science  Centre  grant  SONATA
BIS  no.  2016/22/E/ST2/00013. F.S.  is  funded  by  the Deutsche  Forschungsgemeinschaft  (DFG)  project  number 392578569.  M.C. acknowledges financial support by
the U.S. Department of Energy, Office of Nuclear Physics,
within the framework of the TMD Topical Collaboration,
as  well  as,  by  the  National  Science  Foundation  under Grant  No.  PHY-1714407.  This  research  used  resources of the Oak Ridge Leadership Computing Facility, which is a DOE Office of Science User Facility supported under Contract DE-AC05-00OR22725, J\"{u}lich Supercomputing Centre, Prometheus supercomputer at the Academic Computing Centre Cyfronet
AGH  in  Cracow  (grant  ID quasipdfs),  Okeanos  supercomputer at the Interdisciplinary Centre for Mathematical and Computational Modelling in Warsaw (grant IDs
gb70-17,  ga71-22),  Eagle  supercomputer  at  the  Poznan Supercomputing and Networking Center (grant no. 346).

\vspace*{-0.3cm}
\bibliographystyle{JHEP}
\bibliography{biblio}

\providecommand{\href}[2]{#2}\begingroup\raggedright\begin{thebibliography}{10}

\bibitem{Lin:2017snn}
H.-W. Lin et~al., \emph{{Parton distributions and lattice QCD calculations: a
  community white paper}},  \href{https://arxiv.org/abs/1711.07916}{{\ttfamily
  1711.07916}}.

\bibitem{Ji:2014gla}
X.~Ji, \emph{{Parton Physics from Large-Momentum Effective Field Theory}},
  \href{https://doi.org/10.1007/s11433-014-5492-3}{\emph{Sci. China Phys. Mech.
  Astron.} {\bfseries 57} (2014) 1407}
  [\href{https://arxiv.org/abs/1404.6680}{{\ttfamily 1404.6680}}].

\bibitem{Radyushkin:2017cyf}
A.~V. Radyushkin, \emph{{Quasi-parton distribution functions, momentum
  distributions, and pseudo-parton distribution functions}},
  \href{https://doi.org/10.1103/PhysRevD.96.034025}{\emph{Phys. Rev.}
  {\bfseries D96} (2017) 034025}
  [\href{https://arxiv.org/abs/1705.01488}{{\ttfamily 1705.01488}}].

\bibitem{Orginos:2017kos}
K.~Orginos, A.~Radyushkin, J.~Karpie and S.~Zafeiropoulos, \emph{{Lattice QCD
  exploration of parton pseudo-distribution functions}},
  \href{https://doi.org/10.1103/PhysRevD.96.094503}{\emph{Phys. Rev.}
  {\bfseries D96} (2017) 094503}
  [\href{https://arxiv.org/abs/1706.05373}{{\ttfamily 1706.05373}}].

\bibitem{Ma:2017pxb}
Y.-Q. Ma and J.-W. Qiu, \emph{{Exploring Partonic Structure of Hadrons Using ab
  initio Lattice QCD Calculations}},
  \href{https://doi.org/10.1103/PhysRevLett.120.022003}{\emph{Phys. Rev. Lett.}
  {\bfseries 120} (2018) 022003}
  [\href{https://arxiv.org/abs/1709.03018}{{\ttfamily 1709.03018}}].

\bibitem{Alexandrou:2018pbm}
C.~Alexandrou et~al., \emph{{Light-Cone Parton Distribution Functions from
  Lattice QCD}},
  \href{https://doi.org/10.1103/PhysRevLett.121.112001}{\emph{Phys. Rev. Lett.}
  {\bfseries 121} (2018) 112001}
  [\href{https://arxiv.org/abs/1803.02685}{{\ttfamily 1803.02685}}].

\bibitem{Alexandrou:2018eet}
C.~Alexandrou, K.~Cichy, M.~Constantinou, K.~Jansen, A.~Scapellato and
  F.~Steffens, \emph{{Transversity parton distribution functions from lattice
  QCD}},  \href{https://arxiv.org/abs/1807.00232}{{\ttfamily 1807.00232}}.

\bibitem{Abdel-Rehim:2015pwa}
{\scshape ETM} collaboration, A.~Abdel-Rehim et~al., \emph{{First physics
  results at the physical pion mass from $N_f=2$ Wilson twisted mass fermions
  at maximal twist}},
  \href{https://doi.org/10.1103/PhysRevD.95.094515}{\emph{Phys. Rev.}
  {\bfseries D95} (2017) 094515}
  [\href{https://arxiv.org/abs/1507.05068}{{\ttfamily 1507.05068}}].

\bibitem{Alexandrou:2018egz}
C.~Alexandrou et~al., \emph{{Simulating twisted mass fermions at physical
  light, strange and charm quark masses}},
  \href{https://doi.org/10.1103/PhysRevD.98.054518}{\emph{Phys. Rev.}
  {\bfseries D98} (2018) 054518}
  [\href{https://arxiv.org/abs/1807.00495}{{\ttfamily 1807.00495}}].

\bibitem{Bali:2016lva}
G.~S. Bali, B.~Lang, B.~U. Musch and A.~Schafer, \emph{{Novel quark smearing
  for hadrons with high momenta in lattice QCD}},
  \href{https://doi.org/10.1103/PhysRevD.93.094515}{\emph{Phys. Rev.}
  {\bfseries D93} (2016) 094515}
  [\href{https://arxiv.org/abs/1602.05525}{{\ttfamily 1602.05525}}].

\bibitem{Krzysztof_proc:18}
Alexandrou et~al., \emph{{Light-cone PDFs from lattice QCD}},  in
  \emph{{\pos{PoS(LATTICE2018)095}}}.

\bibitem{Alexandrou:2017huk}
C.~Alexandrou et~al., \emph{{A complete non-perturbative renormalization
  prescription for quasi-PDFs}}, {\emph{Nucl. Phys.} {\bfseries B923} (2017)
  394} [\href{https://arxiv.org/abs/1706.00265}{{\ttfamily 1706.00265}}].

\bibitem{Alexandrou:2015sea}
{\scshape ETM} collaboration, C.~Alexandrou, M.~Constantinou and
  H.~Panagopoulos, \emph{{Renormalization functions for Nf=2 and Nf=4 twisted
  mass fermions}},
  \href{https://doi.org/10.1103/PhysRevD.95.034505}{\emph{Phys. Rev.}
  {\bfseries D95} (2017) 034505}
  [\href{https://arxiv.org/abs/1509.00213}{{\ttfamily 1509.00213}}].

\bibitem{Constantinou:2017sej}
M.~Constantinou and H.~Panagopoulos, \emph{{Perturbative renormalization of
  quasi-parton distribution functions}},
  \href{https://doi.org/10.1103/PhysRevD.96.054506}{\emph{Phys. Rev.}
  {\bfseries D96} (2017) 054506}
  [\href{https://arxiv.org/abs/1705.11193}{{\ttfamily 1705.11193}}].

\bibitem{Alexandrou:2017dzj}
C.~Alexandrou et~al., \emph{{Computation of parton distributions from the
  quasi-PDF approach at the physical point}},
  \href{https://doi.org/10.1051/epjconf/201817514008}{\emph{EPJ Web Conf.}
  {\bfseries 175} (2018) 14008}
  [\href{https://arxiv.org/abs/1710.06408}{{\ttfamily 1710.06408}}].

\bibitem{Chen:2016utp}
J.-W. Chen, S.~D. Cohen, X.~Ji, H.-W. Lin and J.-H. Zhang, \emph{{Nucleon
  Helicity and Transversity Parton Distributions from Lattice QCD}},
  \href{https://doi.org/10.1016/j.nuclphysb.2016.07.033}{\emph{Nucl. Phys.}
  {\bfseries B911} (2016) 246}
  [\href{https://arxiv.org/abs/1603.06664}{{\ttfamily 1603.06664}}].

\bibitem{Ball:2017nwa}
{\scshape NNPDF} collaboration, R.~D. Ball et~al., \emph{{Parton distributions
  from high-precision collider data}},
  \href{https://doi.org/10.1140/epjc/s10052-017-5199-5}{\emph{Eur. Phys. J.}
  {\bfseries C77} (2017) 663}
  [\href{https://arxiv.org/abs/1706.00428}{{\ttfamily 1706.00428}}].

\bibitem{Alekhin:2017kpj}
S.~Alekhin, J.~Blumlein, S.~Moch and R.~Placakyte, \emph{{Parton distribution
  functions, $\alpha_s$, and heavy-quark masses for LHC Run II}},
  \href{https://doi.org/10.1103/PhysRevD.96.014011}{\emph{Phys. Rev.}
  {\bfseries D96} (2017) 014011}
  [\href{https://arxiv.org/abs/1701.05838}{{\ttfamily 1701.05838}}].

\bibitem{Accardi:2016qay}
A.~Accardi et~al., \emph{{Constraints on large-$x$ parton distributions from
  new weak boson production and deep-inelastic scattering data}},
  \href{https://doi.org/10.1103/PhysRevD.93.114017}{\emph{Phys. Rev.}
  {\bfseries D93} (2016) 114017}
  [\href{https://arxiv.org/abs/1602.03154}{{\ttfamily 1602.03154}}].

\bibitem{Nocera:2014gqa}
{\scshape NNPDF} collaboration, E.~R. Nocera et~al., \emph{{A first unbiased
  global determination of polarized PDFs and their uncertainties}},
  \href{https://doi.org/10.1016/j.nuclphysb.2014.08.008}{\emph{Nucl. Phys.}
  {\bfseries B887} (2014) 276}
  [\href{https://arxiv.org/abs/1406.5539}{{\ttfamily 1406.5539}}].

\bibitem{deFlorian:2009vb}
D.~de~Florian, R.~Sassot, M.~Stratmann and W.~Vogelsang, \emph{{Extraction of
  Spin-Dependent Parton Densities and Their Uncertainties}},
  \href{https://doi.org/10.1103/PhysRevD.80.034030}{\emph{Phys. Rev.}
  {\bfseries D80} (2009) 034030}
  [\href{https://arxiv.org/abs/0904.3821}{{\ttfamily 0904.3821}}].

\bibitem{Ethier:2017zbq}
J.~J. Ethier et~al., \emph{{First simultaneous extraction of spin-dependent
  parton distributions and fragmentation functions from a global QCD
  analysis}}, \href{https://doi.org/10.1103/PhysRevLett.119.132001}{\emph{Phys.
  Rev. Lett.} {\bfseries 119} (2017) 132001}
  [\href{https://arxiv.org/abs/1705.05889}{{\ttfamily 1705.05889}}].

\bibitem{Lin:2017stx}
H.-W. Lin et~al., \emph{{First Monte Carlo Global Analysis of Nucleon
  Transversity with Lattice QCD Constraints}},
  \href{https://doi.org/10.1103/PhysRevLett.120.152502}{\emph{Phys. Rev. Lett.}
  {\bfseries 120} (2018) 152502}
  [\href{https://arxiv.org/abs/1710.09858}{{\ttfamily 1710.09858}}].

\end{thebibliography}\endgroup

\end{document}